\documentclass[aps,prl,twocolumn,citeautoscript,footinbib]{revtex4-2}
\usepackage{feynmp}

\usepackage{subcaption}
\usepackage{amsmath}
\usepackage{amssymb}
\usepackage{bbm}
\usepackage{braket}
\usepackage{xcolor}
\usepackage{pifont}
\usepackage{slashed}
\usepackage[mathscr]{euscript}
\usepackage[shortlabels]{enumitem}
\usepackage{tikz-feynman}
\usepackage[papersize={8.5in,11in}]{geometry}

\allowdisplaybreaks

\usepackage{graphicx}
\usepackage{subcaption}
\usepackage[colorlinks=true]{hyperref}
\usepackage{comment}
\usepackage{multirow}

\renewcommand{\Im}{\text{Im}}

\renewcommand{\vec}[1]{\boldsymbol{#1}}
\newcommand{\vn}[1]{|\boldsymbol{#1}|}

\definecolor{wrongultramarine}{rgb}{1,0.5,0}

\newcommand{\rd}{{\rm d}}
\newcommand{\sgn}{{\rm sgn\,}}

\newcommand{\calN}{{\mathcal N}}

\newcommand{\verteq}{\rotatebox{90}{$\,=$}}
\newcommand{\vertdots}{\rotatebox{90}{$\,\dots$}}

\hypersetup{
    bookmarks=true,         
    unicode=false,          
    pdftoolbar=true,        
    pdfmenubar=true,        
    pdffitwindow=false,     
    pdfstartview={FitH},    
    pdfsubject={},   
    pdfcreator={},   
    pdfproducer={}, 
    pdfkeywords={} {} {}, 
    pdfnewwindow=true,      
    colorlinks=true,       
    linkcolor=blue, 
    citecolor=blue,        
    filecolor=magenta,      
    urlcolor=blue           
}

\tikzset{
  mid arrow/.style={postaction={decorate,decoration={
        markings,
        mark=at position .575 with {\arrow[#1]{stealth}}
      }}},
  near arrow/.style={postaction={decorate,decoration={
        markings,
        mark=at position .275 with {\arrow[#1]{stealth}}
      }}},
   far arrow/.style={postaction={decorate,decoration={
        markings,
        mark=at position .800 with {\arrow[#1]{stealth}}
      }}},
   boson/.style={decorate, draw=black,
    decoration={snake,amplitude=1pt, segment length=5pt},
      },
   mid triangle/.style={postaction={decorate,decoration={
        markings,
        mark=at position .575 with {\arrow[#1]{triangle 45}}
      }}}
}

\usepackage{pdfpages} 
\usepackage{pgffor} 
\makeatletter
\AtBeginDocument{\let\LS@rot\@undefined}
\makeatother

\def\supplementfilename{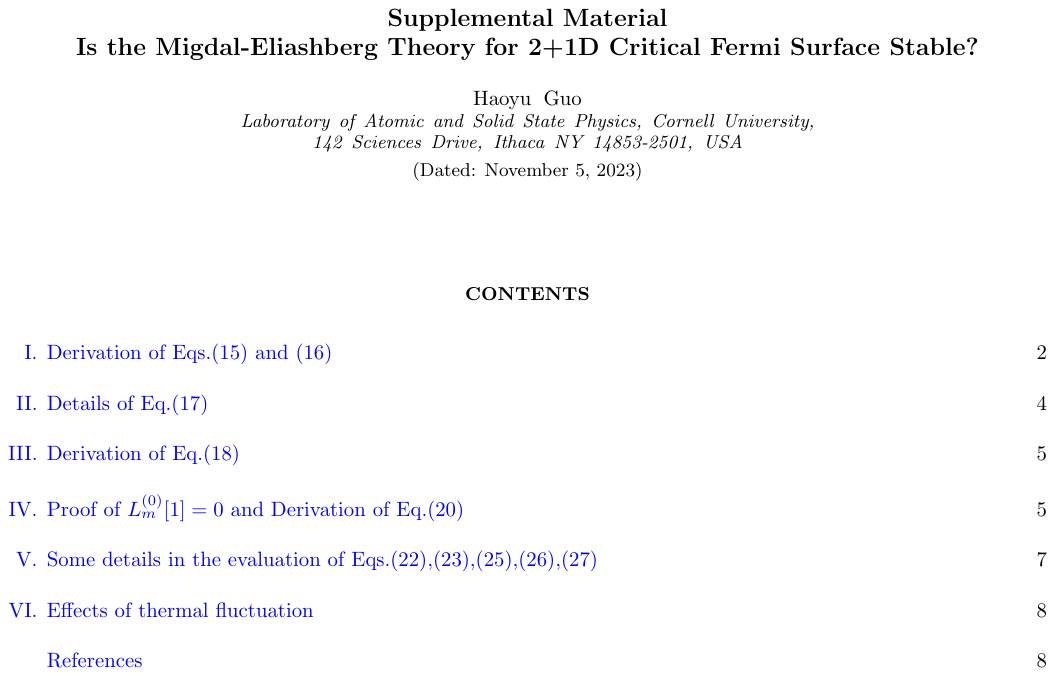}

\pdfximage{\supplementfilename}
\def\numbersupplementpages{\the\pdflastximagepages}

\newif\ifarXiv
\arXivtrue

\begin{document}

\title{Is the Migdal-Eliashberg Theory for 2+1D Critical Fermi Surface Stable?}
\begin{abstract}
    We diagnose the stability of the Migdal-Eliashberg theory for a Fermi surface coupled to a gapless boson in 2+1 dimensions. We provide a scheme for diagonalizing the Bethe-Salpeter ladder when small-angle scattering mediated by the boson plays a dominant role. We found a large number of soft modes which correspond to shape fluctuations of the Fermi surface, and these shape deformations follow a diffusion-like dynamics on the Fermi surface. Surprisingly, the odd-parity deformations of a convex Fermi surface becomes unstable near the non-Fermi liquid regime of the Ising-Nematic quantum critical point and our finding calls for revisit of the Migdal-Eliashberg framework. The implication of the Bethe-Salpeter eigenvalues in transport will be discussed in the companion paper [\href{https://doi.org/10.48550/arXiv.2311.03458}{H.Guo,arXiv:2311.03458}].
\end{abstract}

\author{Haoyu Guo}
\affiliation{Laboratory of Atomic and Solid State Physics, Cornell University,
142 Sciences Drive, Ithaca NY 14853-2501, USA}

\date{\today}

\maketitle
\newpage

\paragraph{Introduction}
    The model of a Fermi surface (FS) coupled to gapless bosonic fluctuations (critical Fermi surface) \cite{SSLee2018} in 2+1 specetime dimensions plays a central role in the study of finite-density fermionic quantum matter \cite{PALee1989,AJMillis1993,JPolchinski1994, BIHalperin1993,YBKim1994a,CNayak1994,SSLee2009,MAMetlitski2010,
    DFMross2010,SSur2014,MAMetlitski2015,SAHartnoll2014,
    AEberlein2017,THolder2015,THolder2015a,ALFitzpatrick2014,
    JADamia2019,JADamia2020,JADamia2021,SPRidgway2015,AAPatel2018b,
    DChowdhury2018c,EGMoon2010,AAbanov2020,YMWu2020,AVChubukov2020,
    XWang2019,AKlein2020,OGrossman2021,DChowdhury2020,IEsterlis2019,
    DHauck2020,YWang2020a,EEAldape2022,AAPatel2017a,AAPatel2019,
    VOganesyan2001,AAPatel2018,AVChubukov2017,DLMaslov2017,
    SLi2023,Iesterlis2021,HGuo2022a,AAPatel2023,HGuo2023a,LVDelacretaz2022a,
    SEHan2023,UMehta2023,DVElse2021a,DVElse2021,ZDShi2022,ZDShi2023,TPark2023}, including the half-filled Landau level, quantum spin liquids, metallic quantum criticality and strange metal.  An analytical framework to the problem which has been deemed historically successful is the Migdal-Eliashberg theory (MET) \cite{PALee1989,BIHalperin1993,YBKim1994a,MAMetlitski2010,
    DFMross2010,EEAldape2022,Iesterlis2021,HGuo2022a,
    ZDShi2022,ZDShi2023,FMarsiglio2020}. MET assumes that the typical boson velocity $v_B$ is much slower than the Fermi velocity $v_F$. This condition can be rewritten in terms of comparing the typical boson momentum $q$ and the fermion energy $\xi_k$, which reads
    \begin{equation}\label{eq:METassump}
      v_F q\gg \xi_k\,.
    \end{equation} Assuming \eqref{eq:METassump}, it can be shown that the vertex corrections can be ignored in self-energy diagrams, and a closed set of Schwinger-Dyson (SD) equations can be derived as
    \begin{equation}\label{eq:MET}
      \begin{split}
         G(i\omega,\vec{k}) & =\frac{1}{i\omega-\xi_{\vec{k}}-\Sigma(i\omega,\vec{k})}\,, \\
         D(i\Omega,\vec{k})  &= \frac{1}{\vn{q}^2+m_b^2-\Pi(i\Omega,\vec{q})}\,, \\
         \Sigma(\tau,\vec{r})  & =g^2 G(\tau,\vec{r})D(\tau,\vec{r})\,, \\
         \Pi(\tau,\vec{r})  &=-g^2 G(\tau,\vec{r})G(-\tau,-\vec{r})\,.
      \end{split}
    \end{equation} Here, the SD equations are written down in the context of a FS $\psi$ coupled to an Ising-Nematic order parameter $\phi$ with Yukawa coupling $g$. The boson mass $m_b^2$ is the tuning parameter for accessing the quantum critical point (QCP). At the QCP $(m_b^2=0)$, the fermionic quasiparticles are destroyed and the system is dubbed a non-Fermi liquid (NFL). $G$ and $D$ are the fermion and the boson Green's functions repsectively and $\Sigma$, $\Pi$ are the corresponding self-energies. $\xi_{\vec{k}}$ is the fermion dispersion measured from the FS. We use units where the boson velocity $v_B=1$ and we drop the $\Omega^2$ bare dynamics of $\phi$ because the self-generated dynamics from the Yukawa coupling is more dominant. Eq.\eqref{eq:MET} has appeared in the literature in the form of random phase approximation (RPA) \cite{PALee1989,BIHalperin1993} or as the saddle point of various large-$N$ expansions \cite{MAMetlitski2010,
    DFMross2010,EEAldape2022,Iesterlis2021,HGuo2022a,
    ZDShi2022,ZDShi2023}. Within this work, we will focus on theories in which the boson dispersion has a minimum at $q=0$, so the entire FS is hot. Throughout our discussion, we ignore umklapp scattering and assume momentum conservation.

    In this letter, we calculate the fluctuation spectrum of the theory \eqref{eq:MET} in the normal state and diagnose its stability in various regimes near the NFL. At first glance, Eq.\eqref{eq:MET} seems to define a sensible theory because the SD equations can be rewritten in terms of convolutions of spectral functions, and then causality and unitarity are automatically satisfied. However, it is unclear whether the theory is stable against collective fluctuations, which are captured by the two-particle-irreducible effective action \cite{DBenedetti2021,JMCornwall1974} or the $G$-$\Sigma$ action \cite{HGuo2022a,YGu2020}:
    \begin{equation}\label{}
      S[\delta G]=\frac{1}{2}\int_{x_1,x_2,x_3,x_4} \delta G(x_2,x_1)K_\text{BS}(x_1,x_2;x_3,x_4)\delta G(x_3,x_4)\,.
    \end{equation} Here $x_i$'s denote spacetime coordinates and $\delta G$ is the fluctuation of $G(x_1,x_2)$ about the saddle point \eqref{eq:MET}. The kernel $K_\text{BS}$ is the generator of the celebrated Bethe-Salpeter (BS) equation. In this letter, we provide a scheme for diagonalizing $K_\text{BS}$ in the critical Fermi surface problem. From this procedure, we find a large number of soft modes which characterize the shape fluctuations of the FS, and surprisingly, half of the soft eigenvalues become negative in the NFL regime, indicating an instability of the MET. In addition to the soft modes, we also found nonzero eigenvalues which is responsible for the dissipative optical conductivity, which will be discussed in the companion paper \cite{prbpaper}.

     \paragraph{The soft modes of the critical FS} The soft modes we found can be intuitively understood as the angular dynamics of a FS. Since the critical FS is strongly coupled to long-wavelength bosons, the typical scattering angle of a fermion is very small. At zeroth order, a particle excess on the FS will stay in a patch centering it for a long period of time, i.e. the local density $n(\theta)$ ($\theta$ labels angle on the FS) is approximately conserved. This reasoning is the foundation for various patch formulations of a FS \cite{SSachdev1999b,SSLee2018,DVElse2021a, ZDShi2022}. Since $n(\theta)$ also parameterizes shape fluctuations of the FS, it also becomes the starting point of recent bosonization studies \cite{LVDelacretaz2022a,UMehta2023,SEHan2023}. Going beyond zeroth order, a particle excess has to move out of its current patch eventually, and statistically it random walks on the FS, so $n(\theta)$ should satisfy a diffusion-like dynamics $\partial_t n=D_k \partial_\theta^{k}n$, and the diffusion constant defines a time scale $\tau\sim D_k^{-1}$ above which the approximate conservation of $n(\theta)$ breaks down. The details of the diffusion dynamics depend on the inversion parity and the FS geometry. For the even-parity modes $n_e(\theta)=n_e(\theta+\pi)$, the diffusion dynamics is generically a regular diffusion dynamics $\partial_t n_e=D_{e}\partial_\theta^2 n_e$ where the diffusion coefficient scales as
     \begin{equation}\label{eq:De}
       D_e\sim\Gamma_\text{sp}\times \braket{q^2}/k_F^2\,,
     \end{equation} where $\Gamma_\text{sp}=\Im \Sigma_R$ is the single-particle scattering rate extracted from the self-energy, and $\sqrt{\braket{q^2}}$ is the typical boson momentum involved in scattering. This result naturally coincides with the random walk picture where the diffusion constant is the walk rate times typical step squared. For the odd-parity modes $n_o(\theta)=n_o(\theta+\pi)$, and a concave FS, the diffusion dynamics is similar to $n_e$. However, for convex or circular FS the diffusion is much slower due to the emergent integrability of the FS \cite{DLMaslov2011,HKPal2012,PJLedwith2019,HGuo2022a}, which arises from the fact that the equation $0=\xi_{\vec{k}}=\xi_{\vec{k}-\vec{q}}$ for a fixed $\vec{q}$ only has a pair of reflection-related solutions on a convex FS. This implies that odd-parity deformations are not relaxed if only shape fluctuations are considered. A proper resolution of the diffusion dynamics requires including fluctuations beyond just shape fluctuations, and the resulting diffusion dynamics is $\partial_t n_o=D_o \partial_\theta^6 n_o$, with
     \begin{equation}\label{eq:Do}
       D_o\sim\Gamma_\text{sp}\times \braket{\xi^2}/(k_Fv_F)^2\times \braket{q^2}/k_F^2\,.
     \end{equation} Here $\sqrt{\braket{\xi^2}}$ is the typical dispersion of a fermion away from FS.
     In the NFL regime and with center-of-mass (CoM) frequency $i\Omega$, we found $\Gamma_\text{sp}\sim \Omega^{2/3}$, $\xi\sim \Omega^{2/3}$, $q\sim \Omega^{1/3}$, and $D_o\propto \Omega^{8/3}$ with positive coefficient. Upon analytical continuation to real time $\Omega\to -i\omega+0$, $D_o$ obtains a negative real part and the theory becomes unstable.

     \paragraph{The Kinetic Operator} We now present our calculation that substantiates the qualitative discussion above. To avoid the thermal effects near the QCP, we will work at zero temperature $T=0$. We first determine the form of the Bethe Salpeter kernel $K_\text{BS}$. This can be done by looking at the $1/N$ expansion of the large-$N$ theories that realize MET as the saddle point, such as the Yukawa-SYK model \cite{Iesterlis2021,HGuo2022a} or the double expansion of small-$(z_b-2)$ and large-$N$  \cite{DFMross2010}. It is not surprising that different approaches yield the same result, because ignoring vertex corrections imply the existence of a Luttinger-Ward (LW) functional whose form is uniquely fixed by MET, and $K_\text{BS}$ is simply the second order expansion of the LW functional, which can be written as three parts:
     \begin{equation}\label{eq:KBS}
       K_\text{BS}=W_\Sigma^{-1}-W_\text{MT}-W_\text{AL}.
     \end{equation}  Here $W_\Sigma$, $W_\text{MT}$ and $W_\text{AL}$ are four-point functions that generate the density-of-states, Maki-Thompson and Aslamazov-Larkin diagrams respectively. They can be conveniently defined in real space as ($\delta$ is spacetime $\delta$-function)
     \begin{equation}\label{}
       W_\Sigma(x_1,x_2;x_3,x_4)=G(x_1,x_3)G(x_4,x_2)\,,
     \end{equation}
     \begin{equation}\label{}
       W_\text{MT}(x_1,x_2;x_3,x_4)=g^2 D(x_3,x_4)\delta(x_1,x_3)\delta(x_2,x_4)\,,
     \end{equation}
     \begin{equation}\label{eq:WAL}
     \begin{split}
       &W_\text{AL}(x_1,x_2;x_3,x_4)=-g^4G(x_1,x_2)G(x_4,x_3)\\
       &\times\left[D(x_1,x_3)D(x_2,x_4)+D(x_1,x_4)D(x_2,x_3)\right]\,.
     \end{split}
     \end{equation} To define the eigenvalue problem  we still need to specify the inner product between two-point functions. Now we require that the conserved quantities (charge and momentum) are exact zero modes of the eigenvalue problem, meaning that the Ward identities should be interpreted as eigenvector equations. Due to spacetime translation symmetry, $K_\text{BS}$ conserves the CoM 3-vector $p$, and from now on we work exclusively on the case $p=(i\Omega,0)$ and consider the retarded branch $\Omega>0$. The remaining relative coordinates are now Fourier transformed to $k=(i\omega,\vec{k})$. The Ward identities \cite{HGuo2022a}  takes the following form:
     \begin{equation}\label{eq:Ward}
       K_\text{BS}[(iG(i\omega+i\Omega/2,\vec{k})-iG(i\omega-i\Omega/2,\vec{k}))\Gamma_\alpha]=\Omega\Gamma_\alpha\,.
     \end{equation} Here $\Gamma_\alpha=1,\vec{k}$ is the charge and momentum vertex, respectively. Therefore, we conjecture that the correct operator $L$ to diagonalize should be
     \begin{equation}\label{}
       L=K_\text{BS}\circ M-\Omega I\,,
     \end{equation} where $\circ$ denotes functional composition,$I$ is the identity operator and $M$ attaches the Green's function factor as in \eqref{eq:Ward} ($F$ is a test two-point function):
     \begin{equation}\label{}
       M[F](i\omega,\vec{k})=(iG(i\omega+i\Omega/2,\vec{k})-iG(i\omega-i\Omega/2,\vec{k}))F(i\omega,\vec{k})\,.
     \end{equation} The inner product $\braket{A|B}$ can be determined from the condition that $L$ remains symmetric, which is
     \begin{equation}\label{eq:innerprod}
     \begin{split}
       &\braket{A|B}=\int\frac{\rd\omega\rd^2\vec{k}}{(2\pi)^3}A(i\omega,\vec{k})\\
       &\times(iG(i\omega+i\Omega/2,\vec{k})-iG(i\omega-i\Omega/2,\vec{k}))
       B(i\omega,\vec{k})\,.
     \end{split}
     \end{equation}
     The eigenvalue problem we want to solve now becomes
     \begin{equation}\label{}
       L\ket{F}=\lambda\ket{F}\,,
     \end{equation} where $\lambda$ is the eigenvalue.
     We refer to $L$ as the kinetic operator, because the integrals that define $L$ resemble those that appear in the quantum Boltzmann equation (QBE) \cite{GDMahan2000} and its eigenvalues can be interpreted as from a collision integral. 

     The inner product \eqref{eq:innerprod} simplifies in the good metal limit $k_F v_F\gg \Sigma$ which we assume in our problem. We choose to perform the momentum integral first and the frequency integral next. This operation does not affect the eigenvalue of $L$ and introduces a contact term error when calculating certain observables such as density correlator, but it can be easily corrected by comparing with free Fermi gas. Next, we assume the functions we are interested in are regular functions in $\xi_{\vec{k}}$. Then the integral over $\xi_{\vec{k}}$ can be done with contour method, and at low-energy the dominant contribution is from the poles of the Green's functions in \eqref{eq:innerprod}. This implies that the integral is only nonzero when $-\Omega/2<\omega<\Omega/2$, so the frequency domain becomes finite.

     \paragraph{Hierarchy of the Kinetic Operator} We now restrict to circular FS and utilize the rotation symmetry to consider functions with angular harmonics $e^{im\theta_k}$ and the corresponding block $L_m$ of $L$. $L_m=L_{\text{DOS+MT},m}+L_{\text{AL},m}$ can be written as a functional acting only on the $(i\omega,\xi)$ domain, which reads \cite{Supp}

     \begin{widetext}
     \begin{equation}\label{eq:LMTT}
        \begin{split}
          L_{\text{MT+DOS},m}[F]&(i\omega,\xi)=g^2\int_{-\infty}^{\infty}\frac{\rd \omega'}{2\pi}\frac{\calN \rd \xi'}{2\pi}\int_0^{\infty}\vn{q}\rd\vn{q}J(\vn{k},\vn{k'},\vn{q})D(\vn{q},i\omega-i\omega')\\
          &\times 2\left[iG(i\omega'+i\Omega/2,\xi')-iG(i\omega'-i\Omega/2,\xi')\right]\left[F(i\omega,\xi)-F(i\omega',\xi')T_m\left(\frac{\vn{k}^2+\vn{k'}^2-\vn{q}^2}{2\vn{k}\vn{k'}}\right)\right]\,.
        \end{split}
        \end{equation}
        \vspace{-2.0em}
        \begin{equation}\label{eq:LALT}
        \begin{split}
        L_{\text{AL},m}[F](i\omega_1,\xi_1)&=\frac{g^4}{2}(2\pi)^2\int_{-\infty}^{\infty}\calN^3\frac{\rd \nu}{2\pi}\frac{\rd \omega_2}{2\pi}\frac{\rd \xi_2}{2\pi}\frac{\rd \xi'}{2\pi}\frac{\rd \xi''}{2\pi}\int_0^\infty \vn{q}\rd \vn{q} J(\vn{k_1},\vn{k'},\vn{q}) J(\vn{k_2},\vn{k''},\vn{q})\\
        &\times D(\vn{q},i\nu+i\Omega/2)D(\vn{q},i\nu-i\Omega/2)\times 4T_m\left(\frac{\vn{q}^2+\vn{k_1}^2-\vn{k'}^2}{2 \vn{q} \vn{k_1}}\right)T_m\left(\frac{\vn{q}^2+\vn{k_2}^2-\vn{k''}^2}{2 \vn{q} \vn{k_2}}\right)\\
        &\times \left[G(i\omega_1-i\nu,\xi')+(-1)^mG(i\omega_1+i\nu,\xi')\right]\left[G(i\omega_2-i\nu,\xi'')+(-1)^mG(i\omega_2+i\nu,\xi'')\right]\\
        &\times i(G(i\omega_2+i\Omega/2,\xi_2)-G(i\omega_2-i\Omega/2,\xi_2))F(i\omega_2,\xi_2)\,.
        \end{split}
        \end{equation}
     \end{widetext} Here, $\calN=k_F/(2\pi v_F)$ is the density of states near the FS. The fermionic momenta $\vn{k}$ are related to $\xi$ with the same label (primed or subscripted) by $\vn{k}=k_F+\xi/v_F$. $J(\vn{k},\vn{k'},\vn{q})=2/({\sqrt{(\vn{k}+\vn{k'})^2-\vn{q}^2}\sqrt{\vn{q}^2-(\vn{k}-\vn{k'})^2}})$ is the Jacobian from angular integration. $T_m$ is the Chebyshev polynomial $T_m(\cos\theta)=\cos m\theta$. Near the QCP, both $\vn{q}$ and $\xi$'s are small and can be expanded as the following: Within the kinematic regime where MET assumption \eqref{eq:METassump} holds, the Jacobian should be approximated by $J=1/(k_F\vn{q})$ and not expanded. The remaining $\xi$ and $\vn{q}$ dependence occur through the Chebyshev polynomials $T_m$, and we expand them first in $\xi$'s and then in $\vn{q}^2$ in accordance with \eqref{eq:METassump}. Following this scheme, we obtain a hierarchy of $L_m$ as
     \begin{alignat}{4}\label{eq:Lexpand}
       L_m=&~L^{(0)}_m\quad+\quad&&~L^{(1)}_m\quad+\quad&&~L^{(2)}_m+\dots \nonumber\\[-2pt]
           &~~\verteq  &&~~\verteq  &&~~\verteq \nonumber\\[-4pt]
           &\delta_q^{0}L_m^{(0)}&&\delta_q^{0}L_m^{(1)}&&\delta_q^{0}L_m^{(2)}\\[-4pt]
           &~~+&&~~+&&~~+\nonumber\\[-4pt]
           &\delta_q^{1}L_m^{(0)}&&\delta_q^{1}L_m^{(1)}&&\delta_q^{1}L_m^{(2)}\nonumber\\[-4pt]
           &~~+&&~~+&&~~+\nonumber\\[-5pt]
           &~~~\vertdots&&~~~\vertdots&&~~~\vertdots\nonumber
     \end{alignat} The horizontal direction is the expansion in $\xi/(k_F v_F)$. The first term $L_m^{(0)}$ contains only shape fluctuations of the FS and going to the right we include energy fluctuations. The vertical direction is the expansion in $\vn{q}^2/k_F^2$, where the zeroth order term $\delta_q^0$ only contains forward scattering and the first order term $\delta_q^1$ contains small angle scattering at the order $\mathcal{O}(\vn{q}^2/k_F^2)$. The soft modes of $L_m$ can then be studied in a perturbative fashion order by order. The scheme we propose here generalizes the approach in Ref.\cite{PJLedwith2019} for FL to NFL by incorporating the idea of Prange-Kadanoff reduction \cite{REPrange1964,HGuo2022a}.
     The zeroth order term $\delta_q^0 L_m^{(0)}$  takes the form of Eq.\eqref{eq:LMTT} with $T_m$ set to 1. For nonzero eigenvalues of $L_m$, it is sufficient to compute it using $\delta_q^0 L_m^{(0)}$ as the corrections from expanding in $\xi$ or $\vn{q}$ are subdominant in scaling. We defer the discussion of the nonzero eigenvalues to the companion paper \cite{prbpaper} and in this letter we focus on the soft modes.

    At the order of $\delta_q^0 L_m^{(0)}$, there is an obvious zero mode given by the constant function $F(i\omega)=1$ for every $m$, which is the Fourier transform of the local density $n(\theta)$. However, only the density mode $m=0$ and the momentum mode $m=1$ are exactly conserved, and all the higher harmonics $m$ should relax eventually but at a slower rate. To resolve the eigenvalues of these soft modes, we should apply perturbation theory using the higher order terms in Eq.\eqref{eq:Lexpand}.

    \paragraph{Even-$m$ soft modes} Because of MET assumption \eqref{eq:METassump}, we should first apply perturbations in the vertical directions of \eqref{eq:Lexpand}, i.e. including effects of small-angle scattering $\delta_q^1 L_m^{(0)}$. It turns out that this is enough to resolve the eigenvalues of the even-$m$ soft modes. The result is given by a first perturbation theory \cite{Supp}
    \begin{equation}\label{eq:lambda_even}
    \begin{split}
      &\lambda_m^{\text{even}}(i\Omega)=\frac{\braket{1|\delta_q^1 L_m^{(0)}|1}}{\braket{1|1}}=\frac{2g^2\calN}{\Omega/(2\pi)}\int_{-\Omega/2}^{\Omega/2}\frac{\rd\omega\rd \omega'}{(2\pi)^2}\\
      &\times\int_0^\infty\frac{\rd\vn{q}}{k_F}D(\vn{q},i\omega-i\omega')\frac{m^2\vn{q}^2}{2k_F^2}\,.
    \end{split}
    \end{equation} Here $\ket{1}$ means the constant function $F(i\omega)=1$. The scaling of $\lambda_m^{\text{even}}$ can be naturally read out to be Eq.\eqref{eq:De} by noting that $L_m^{(0)}$ scales the same way as the self-energy and that $\delta_q^1$ contributes a factor of $\vn{q}^2/k_F^2$. The $m^2$ dependence comes from expanding the Chebyshev polynomials and translates to a regular diffusion. As a sanity check $\lambda_0^{\text{even}}=0$ in accordance of charge conservation.

    \paragraph{Odd-$m$ soft modes} The analysis of  odd-$m$ soft modes is more complicated. Due to the kinematic constraint or emergent integrability of a circular FS \cite{DLMaslov2011,HKPal2012,PJLedwith2019,HGuo2022a}, $L_m^{(0)}[1]=0$ to all order in $\vn{q}$. Algebraically, this can be shown by using Eq.\eqref{eq:MET} to transform Eq.\eqref{eq:LALT} when it acts on the constant function and show that it cancels Eq.\eqref{eq:LMTT} \cite{Supp}. Therefore, the perturbation calculation of the eigenvalue must involve expansion in $\xi$. Because the boson $\phi$ is real, there is a particle-hole symmetry $(\omega,\xi)\to(-\omega,-\xi)$ near the FS, under which $L_m^{(0)}$ is even and $L_m^{(1)}$ is odd, so first order perturbation $\braket{1|L_m^{(1)}|1}$ vanishes identically. A nonzero answer requires perturbing second order in $\xi$ and first order in $\vn{q}^2$ :
    \begin{equation}\label{eq:lambda_odd0}
      \lambda_m^{\text{odd}}=\frac{\delta_q^1}{\braket{1|1}}\left[\braket{1|L_m^{(2)}|1}-\braket{1|L_m^{(1)}\frac{1}{L_m^{(0)}}L_m^{(1)}|1}\right]\,.
    \end{equation} The reason to include first order in $\vn{q}^2$ perturbation is because the zeroth order forward angle scattering does not lead to relaxation. The functional inverse in Eq.\eqref{eq:lambda_odd} can be evaluated analytically, with the result \cite{Supp}
    \begin{equation}\label{eq:lambda_odd}
    \begin{split}
      &\lambda_m^\text{odd}=\frac{2g^2\calN}{\Omega/(2\pi)} \int_{-\Omega/2}^{\Omega/2}\frac{\rd \omega\rd \omega'}{(2\pi)^2}\int_{-\infty}^\infty\frac{\rd \xi\rd \xi'}{(2\pi)^2}\int_0^{\infty}\frac{\rd\vn{q}}{k_F}\\
      &\times D(\vn{q},i\omega-i\omega')\left[iG(i\omega+i\Omega/2,\xi)-iG(i\omega-i\Omega/2,\xi)\right]\\
      &\times\left[iG(i\omega'+i\Omega/2,\xi')-iG(i\omega'-i\Omega/2,\xi')\right]
      \\&\times\frac{\vn{q}^2}{k_F^2} \frac{m^2(m^2-1)^2(\xi+\xi')^2}{8k_F^2 v_F^2}\,.
    \end{split}
    \end{equation} The scaling of the various factors coincide with Eq.\eqref{eq:Do}. When $m\gg 1$, $\lambda_m^{\text{odd}}\propto m^6$ indicating an anomalous diffusion as advertised earlier. Also, $\lambda_1^\text{odd}=0$ in agreement with momentum conservation.

    \paragraph{Application to the Ising-Nematic QCP} We now apply our results to the phase diagram near the Ising-Nematic QCP. Near the QCP the boson self-energy is given by Landau damping $\Pi(i\Omega,q)=-\gamma|\Omega|/\vn{q}$ with $\gamma=g^2\calN/v_F$. We first consider the Fermi-liquid (FL) region where the boson has a mass $m_b\ll k_F$. The system is in the FL phase when the typical frequency satisfies $\omega\ll \omega_\text{FL}=m_b^3/\gamma$. The fermion self-energy is then
    \begin{equation}\label{}
        \Sigma(i\omega)=(-i\omega)c_f'\left(\pi+\frac{|\omega|}{\omega_{\text{FL}}}\ln\left(\frac{|\omega|}{\omega_{\text{FL}}}\right)\right),
    \end{equation} where $c_f'=g^2\calN/(2\pi k_F m_b)$. The first term is due to elastic scattering and the second term is due to Landau damping \cite{AVChubukov2003}. Evaluating \eqref{eq:lambda_even} and \eqref{eq:lambda_odd}, we obtain \cite{Supp}
    \begin{equation}\label{eq:lambda_even_C}
     \lambda_m^\text{even}=-c_f' \frac{m^2 m_b^2}{2k_F^2}\left(\pi\Omega+\frac{\Omega^2}{3\omega_{\text{FL}}}\right)\,,
    \end{equation}
    \begin{equation}\label{eq:lambda_odd_C}
  \lambda_m^{\text{odd}}=c_f'\Omega \frac{m^2(m^2-1)^2 m_b^2}{120 k_F^2} \frac{\Omega^2}{k_F^2 v_F^2}\left[10\pi+3\frac{\Omega}{\omega_\text{FL}}\right](1+\pi c_f')^2\,.
\end{equation} As a stability test, we analytically continue $i\Omega\to \omega+i0$, both \eqref{eq:lambda_even_C} and \eqref{eq:lambda_odd_C} has a positive real part, meaning that FL is stable.

    We now move to the QCP where $m_b^2=0$. The boson has dynamical exponent $z_b=3$ and the fermion self-energy now becomes \cite{Iesterlis2021, HGuo2022a}
    \begin{equation}\label{}
      \Sigma(i\omega)=-ic_f|\omega|^{2/3}\sgn \omega\,,\quad c_f=\frac{g^2}{2\sqrt{3}\pi v_F \gamma^{1/3}}.
    \end{equation} This defines a scale $\omega_0=c_f^3\propto g^4/(k_F v_F)$. When $\omega\gg \omega_0$, $|\Sigma|\ll \omega$ and the system is a perturbative NFL meaning that although the self-energy dominates in scaling when $\omega\to 0$, its actual magnitude is smaller than the bare $i\omega$ term. In this regime, we have \cite{Supp}
    \begin{equation}\label{eq:lambda_even_AB}
      \lambda_m^\text{even}=-\frac{3\sqrt{3}m^2}{14}\frac{\left(c_f\Omega^{2/3}\right)^2}{k_F v_F}\,.
    \end{equation}
    \begin{equation}\label{eq:lambda_odd_B}
      \lambda_m^\text{odd}=\frac{c_f^2 \Omega^{10/3}}{k_F^3v_F^3} m^2(m^2-1)^2\times \frac{249\sqrt{3}}{7280}\,.
    \end{equation} Both eigenvalues are stable after continuing to real frequency.

    Finally, the NFL regime is accessed when $\omega\ll \omega_0$ and $|\Sigma|\gg \omega$. In this regime \eqref{eq:lambda_even_AB} still holds, and for the odd-$m$ soft modes we have \cite{Supp}
    \begin{equation}\label{eq:lambda_odd_A}
      \lambda_m^{\text{odd}}=\frac{c_f^4 \Omega^{8/3}}{k_F^3v_F^3} m^2(m^2-1)^2\times 0.077056\,.
    \end{equation} 
    However, Eq.\eqref{eq:lambda_odd_A} has a negative real part after continued to real frequency, indicating an instability.

    \paragraph{Discussion} Our result applies when $m<m_c$, where $m_c^2=\min(k_F^2/\braket{\vn{q}^2},k_F v_F/\braket{\xi})$. This bound arises because the $m$-dependence originates from expanding the Chebyshev polynomials which satisfy $|T_m|\leq 1$. When $m>m_c$, the $m^2 v_F\vn{q}$ and $m^2\xi$ should be replaced by $k_F v_F$ and the soft mode eigenvalues become comparable to the self-energy.

    The instability we discovered relies on the emergent integrability of 2+1D circular or convex FS, and for non-circular FS, $m$ should be interpreted as the Laplacian eigenvalue on the FS. For concave FS in 2D and FS in 3D, the cancellation does not happen and we expect $\lambda_m^\text{odd}\sim \lambda_m^\text{even}$ so the instability does not appear. The instability also depends on the $z=3$ QCP at $T=0$ and thermal fluctuation does not lead to instability \cite{Supp}.



    The results we have found are not contradictory with previous numerical studies including Monte-Carlo simulations \cite{AKlein2020,XYXu2020} and numerical solution of Eq.\eqref{eq:MET} \cite{Iesterlis2021} because they only accessed the perturbative NFL regime of the $z_b=3$ criticality.

    The calculation we have presented is essentially a phase-space counting argument and does not care about the form factor of the Yukawa coupling (it can be shown that the form factors are squared). Therefore, our results also apply to the case of a Fermi surface coupled to U(1) gauge field. The instability we have found is distinct from other symmetry-breaking instabilities such as CDW, pairing or nematicity because it is independent of whether the boson mediates attraction/repultion and it happens in odd-parity channel. We also note that the instability we found does compete with the symmetry-breaking instabilities. For example in the Ising-Nematic QCP the strongest pairing $T_c$ is found to be comparable with $\omega_0$ \cite{YWang2016}.


    We can redo the calculation for general boson dynamical exponent $z_b$ by changing the propagator $D^{-1}=\vn{q}^{z_b-1}+\gamma|\Omega|/\vn{q}$. For $2\leq z_b\leq3$, the perturbative NFL has $\lambda_m^\text{even}\propto -\Omega^{4/z_b}$ and $\lambda_m^\text{odd}\propto \Omega^{2+4/z_b}$ and it is always stable. The NFL regime has $\lambda_m^\text{odd}\propto \Omega^{8/z_b}$ and the stability condition is $2<z_b<8/3$.  A possible resolution of the instability is that $z_b$ receives an $\mathcal{O}(1)$ correction \cite{Shi_unpub,JRech2006,SPRidgway2015} that falls into the stability bound.

    \paragraph{Conclusion} By calculating the soft eigenvalues of the Bethe-Salpeter kernel within the Migdal-Eliashberg theory, we found the odd parity deformations of a circular FS is an instability of the $z_b=3$ critical Fermi surface in the non-Fermi liquid regime at zero temperature. Our finding calls for revisits of the Migdal-Eliashberg framework.

    \begin{acknowledgments} We thank Dmitrii L. Maslov, Andrey V. Chubukov and Alex Levchenko for the inspiring discussions  that initiated this work at the KITP program ``Quantum Materials With And Without Quasiparticles'' . KITP is supported in part by the National Science Foundation under Grants No. NSF PHY-1748958 and PHY-2309135. We thank Debanjan Chowdhury, Zhengyan Darius Shi, Hart Goldman, Senthil Todadri, Leonid Levitov, J\"org Schmalian, Aavishkar A. Patel, Ilya Esterlis and Subir Sachdev for helpful discussions. Haoyu Guo is supported by the Bethe-Wilkins-KIC postdoctoral fellowship at Cornell University.
    \end{acknowledgments}
\bibliography{NFL,supp}

\ifarXiv
    

\section*{Supplementary material}

# I. DERIVATION OF EQS.(15) AND (16)

In this section, we provide derivation for Eqs. (15) and (16) of the main text.

We start by rewriting the kernels specified in Eqs.(7)-(9) of the main text. Because of spacetime translational invariance of the problem, any two-point function can be fourier transformed using a center-of-mass (CoM) 3-momentum $p$ and relative 3-momentum $k$:

$$F(k; p) = \int \mathrm{d}^3 x_1 \mathrm{d}^3 x_2 F(x_1, x_2) \exp(-ip \cdot (x_1 + x_2)/2 - ik \cdot (x_1 - x_2)). \tag{1.1}$$

Since $p$ is conserved by the Bethe-Salpeter Kernel $K_{\mathrm{BS}}$, we will write $F(k)$ instead of $F(k; p)$ for clarity.

The Bethe-Salpeter kernel is then generated by three sets of kernels [1]

$$K_{\mathrm{BS}} = W_{\Sigma}^{-1} - W_{\mathrm{MT}} - W_{\mathrm{AL}}. \tag{1.2}$$

$W_{\Sigma}$ generates the density-of-states Feynmann diagrams. It is diagonal in the momentum space

$$W_{\Sigma}[F](k) = G(k + p/2)G(k - p/2)F(k). \tag{1.3}$$

$W_{\mathrm{MT}}$ generates the Maki-Thompson diagram

$$W_{\mathrm{MT}}[F](k) = g^2 \int \frac{\mathrm{d}^3 k'}{(2\pi)^3} D(k - k')F(k'). \tag{1.4}$$

$W_{\mathrm{AL}}$ generates the Aslamazov-Larkin diagrams

$$W_{\mathrm{AL}}[F](k_1; p) = -g^4 \int \frac{\mathrm{d}^3 q \mathrm{d}^3 k_2}{(2\pi)^6} G(k_1 - q) \left( G(k_2 - q) + G(k_2 + q) \right) D(q + p/2)D(q - p/2) \\ \times F(k_2), \tag{1.5}$$

Next we express the kinetic operator $L$ in the momentum space. From now on we set $p = (i\Omega, 0)$. By definition $L = K_{\mathrm{BS}} \circ M$, where $M$ is a functional diagonal in the momentum space:

$$M(k) = iG(k + i\Omega/2) - iG(k - i\Omega/2). \tag{1.6}$$

Here we added a 3-vector $k$ to a scalar $i\Omega/2$, and it is understood that only the frequency component is added.

We can combine $W_{\Sigma}$ and $M$ to write

$$W_{\Sigma}^{-1}M = iG^{-1}(k - i\Omega/2) - iG^{-1}(k + i\Omega/2) = \Omega + i\Sigma(k + i\Omega/2) - i\Sigma(k - i\Omega/2). \tag{1.7}$$

The $\Omega$ term will be cancelled by the definition of $L$ (Eq.(11) of the maintext). The self-energy can be rewritten using the Eliashberg equation

$$\Sigma(k \pm i\Omega/2) = g^2 \int \frac{\mathrm{d}^3 k'}{(2\pi)^3} D(k - k')G(k' \pm i\Omega/2). \tag{1.8}$$

We therefore combine $W_{\Sigma}$ and $W_{\mathrm{MT}}$ to write

$$L_{\mathrm{MT+DOS}}[F] = (W_{\Sigma}^{-1} - W_{\mathrm{MT}}) \circ M[F] = g^2 \int \frac{\mathrm{d}^3 k'}{(2\pi)^3} D(k - k') \left[ iG(k' + i\Omega/2) - iG(k' - i\Omega/2) \right] \left[ F(k) - F(k') \right]. \tag{1.9}$$

The remaining Aslamazov-Larkin part is

$$L_{\mathrm{AL}}[F](k_1) = g^4 \int \frac{\mathrm{d}^3 q \mathrm{d}^3 k_2}{(2\pi)^6} G(k_1 - q) \left( G(k_2 - q) + G(k_2 + q) \right) D(q + i\Omega/2)D(q - i\Omega/2) \\ \times \left[ iG(k_2 + i\Omega/2) - iG(k_2 - i\Omega/2) \right] F(k_2). \tag{1.10}$$

We now switch to the angular harmonical basis by setting

$$F(k) = F(i\omega, \xi_k)e^{im\theta_k}$$

, where $\xi_k$ is the dispersion of the 2-momentum $\boldsymbol{k}$ measured from the FS, and $\theta_k$ denotes its direction.

The $m$-th harmonic component of $L$ is then defined to be

$$L_m[F](i\omega, \xi) = \int_0^{2\pi} \frac{\mathrm{d}\theta_k}{2\pi} e^{-im\theta} L[Fe^{im\theta_k}](i\omega, \xi, \theta) \,. \tag{1.11}$$

Eq.(1.9) becomes

$$L_{\text{MT+DOS},m}[F](i\omega, \xi_k) = \int_0^{2\pi} \frac{\mathrm{d}\theta_k}{2\pi} \frac{\mathrm{d}\theta_k'}{2\pi} \int_{-\infty}^{\infty} \frac{\mathcal{N}\mathrm{d}\xi_k'}{2\pi} \int_{-\infty}^{\infty} \frac{\mathrm{d}\omega'}{2\pi} D(\boldsymbol{k}-\boldsymbol{k}', i\omega - i\omega') \left[ iG(i\omega' + i\Omega/2, \xi_k') - iG(i\omega' - i\Omega/2, \xi_k') \right]$$
$$\times \left[ F(i\omega, \xi_k) - e^{im(\theta_k' - \theta_k)} F(i\omega', \xi_k') \right] \tag{1.12}$$

Here $\boldsymbol{k}' = (\xi_k', \theta_k')$ and $\boldsymbol{k} = (\xi_k, \theta_k)$. We compute the angular integrals by introducing the bosonic momentum $\boldsymbol{q}$ and inserting the identity $1 = \int \mathrm{d}^2\boldsymbol{q}\delta(\boldsymbol{q} = \boldsymbol{k} - \boldsymbol{k}')$. The integral over $\theta_k$ and $\theta_k'$ can then be calculated by solving the $\delta$-function with $|\boldsymbol{k}|$ and $|\boldsymbol{k}'|$ fixed. For a circular Fermi surface, there are two solutions with

$$\theta_k - \theta_k' = \pm \cos^{-1} \frac{|\boldsymbol{k}|^2 + |\boldsymbol{k}'|^2 - |\boldsymbol{q}|^2}{2|\boldsymbol{k}||\boldsymbol{k}'|} \,. \tag{1.13}$$

The Jacobian associated with $\delta$ function is

$$J(|\boldsymbol{k}|, |\boldsymbol{k}'|, |\boldsymbol{q}|) = \frac{2}{\sqrt{(|\boldsymbol{k}| + |\boldsymbol{k}'|)^2 - |\boldsymbol{q}|^2}\sqrt{|\boldsymbol{q}|^2 - (|\boldsymbol{k}| - |\boldsymbol{k}'|)^2}} \,, \tag{1.14}$$

which is inverse proportional to the area of the triangle formed by $\boldsymbol{k}, \boldsymbol{k}', \boldsymbol{q}$. Finally using the definition of Chebyshev polynomials $T_m(\cos\theta) = \cos m\theta$, we arrive at Eq.(15) of the main text.

$$L_{\text{MT+DOS},m}[F](i\omega, \xi) = g^2 \int \frac{\mathrm{d}\omega'}{2\pi} \int \frac{\mathcal{N}\mathrm{d}\xi'}{2\pi} \int |\boldsymbol{q}|\mathrm{d}|\boldsymbol{q}| J(|\boldsymbol{k}|, |\boldsymbol{k}'|, |\boldsymbol{q}|) D(|\boldsymbol{q}|, i\omega - i\omega')$$
$$\times 2 \left[ iG(i\omega' + i\Omega/2, \xi') - iG(i\omega' - i\Omega/2, \xi') \right] \left[ F(i\omega, \xi) - F(i\omega', \xi') T_m \left( \frac{|\boldsymbol{k}|^2 + |\boldsymbol{k}'|^2 - |\boldsymbol{q}|^2}{2|\boldsymbol{k}||\boldsymbol{k}'|} \right) \right] \,. \tag{1.15}$$

Similar manipulations can also be applied to Eq.(1.10). We introduce dummy momentum variables $\boldsymbol{k}' = \boldsymbol{k_1} - \boldsymbol{q}$ and $\boldsymbol{k}'' = \boldsymbol{k_2} \mp \boldsymbol{q}$. Eq.(1.10) becomes

$$L_{\text{AL},m}[F](i\omega_1, \xi_1) = g^4 (2\pi)^4 \int \frac{\mathrm{d}\theta_1}{2\pi} \int \frac{\mathrm{d}^2\boldsymbol{q}\mathrm{d}\nu}{(2\pi)^3} \int \frac{\mathcal{N}\mathrm{d}\xi_2\mathrm{d}\theta_2}{2\pi} \int \frac{\mathcal{N}\mathrm{d}\xi'\mathrm{d}\theta'}{2\pi} \frac{\mathcal{N}\mathrm{d}\xi''\mathrm{d}\theta''}{2\pi} D(\boldsymbol{q}, i\nu + i\Omega/2) D(\boldsymbol{q}, i\nu - i\Omega/2)$$
$$G(i\omega_1 - i\nu, \xi')\delta(\boldsymbol{q} = \boldsymbol{k_1} - \boldsymbol{k}') \left[ G(i\omega_2 - i\nu, \xi'')\delta(\boldsymbol{q} = \boldsymbol{k_2} - \boldsymbol{k}'') + G(i\omega_2 + i\nu)\delta(\boldsymbol{q} = \boldsymbol{k}'' - \boldsymbol{k_2}) \right]$$
$$\times \left[ iG(i\omega_2 + i\Omega/2, \xi_2) - iG(i\omega_2 - i\Omega/2, \xi_2) \right] F(i\omega_2, \xi_2)e^{im(\theta_2 - \theta_1)} \,. \tag{1.16}$$

Here $\boldsymbol{k_1} = (\xi_1, \theta_1)$, $\boldsymbol{k_2} = (\xi_2, \theta_2)$, $\boldsymbol{k}' = (\xi', \theta')$ and $\boldsymbol{k}'' = (\xi'', \theta'')$. Next we can compute the integrals over $\theta_1, \theta_2, \theta', \theta''$ by solving the $\delta$-functions. There are two solutions $\pm\bar\theta_1$ and $\pm\bar\theta_2$ for $\delta(\boldsymbol{q} = \boldsymbol{k_1} - \boldsymbol{k}')$ and $\delta(\boldsymbol{q} = \boldsymbol{k_2} - \boldsymbol{k}'')$ respectively, where $\bar\theta_1$ and $\bar\theta_2$ represent the angles of $\boldsymbol{k_1}$ and $\boldsymbol{k_2}$ relative to $\boldsymbol{q}$ respectively, under the constrain of the $\delta$-functions. The solution of $\delta(\boldsymbol{q} = \boldsymbol{k}'' - \boldsymbol{k_2})$ is $\pi \pm \bar\theta_2$. The cosines of $\bar\theta_1$ and $\bar\theta_2$ are

$$\cos\bar\theta_1 = \frac{|\boldsymbol{q}|^2 + |\boldsymbol{k_1}|^2 - |\boldsymbol{k}'|^2}{2|\boldsymbol{q}||\boldsymbol{k_1}|} \,, \tag{1.17}$$

$$\cos\bar\theta_2 = \frac{|\boldsymbol{q}|^2 + |\boldsymbol{k_2}|^2 - |\boldsymbol{k}''|^2}{2|\boldsymbol{q}||\boldsymbol{k_2}|} \,. \tag{1.18}$$

After symmetrizing the integral between $\boldsymbol{k}'$ and $\boldsymbol{k}''$, we obtain Eq.(16) of the main text:

$$
\begin{aligned}
L_{\mathrm{AL},m}[F](i\omega_1,\xi_1) = \frac{g^4}{2}(2\pi)^2 \int \mathcal{N}^3 \frac{\mathrm{d}\nu}{2\pi}\frac{\mathrm{d}\omega_2}{2\pi}\frac{\mathrm{d}\xi_2}{2\pi}\frac{\mathrm{d}\xi'}{2\pi}\frac{\mathrm{d}\xi''}{2\pi} \int |\boldsymbol{q}|\mathrm{d}|\boldsymbol{q}| J(|\boldsymbol{k_1}|,|\boldsymbol{k}'|,|\boldsymbol{q}|)J(|\boldsymbol{k_2}|,|\boldsymbol{k}''|,|\boldsymbol{q}|) \\
\times D(|\boldsymbol{q}|,i\nu+i\Omega/2)D(|\boldsymbol{q}|,i\nu-i\Omega/2)\times 4T_m\left(\frac{|\boldsymbol{q}|^2+|\boldsymbol{k_1}|^2-|\boldsymbol{k}'|^2}{2|\boldsymbol{q}||\boldsymbol{k_1}|}\right)T_m\left(\frac{|\boldsymbol{q}|^2+|\boldsymbol{k_2}|^2-|\boldsymbol{k}''|^2}{2|\boldsymbol{q}||\boldsymbol{k_2}|}\right) \\
\times \left[G(i\omega_1-i\nu,\xi')+(-1)^m G(i\omega_1+i\nu,\xi')\right]\left[G(i\omega_2-i\nu,\xi'')+(-1)^m G(i\omega_2+i\nu,\xi'')\right] \\
\times i(G(i\omega_2+i\Omega/2,\xi_2)-G(i\omega_2-i\Omega/2,\xi_2))F(i\omega_2,\xi_2)\,.
\end{aligned}
$$

$$(1.19)$$

## II. DETAILS OF EQ.(17)

We now provide details on the hierarchy described in Eq.(17) of the main text. We express the fermionic momentum $|\boldsymbol{k}|$ ins terms of the dispersion $\xi_k$ with the expansion

$$
|\boldsymbol{k}| = k_F + \frac{\xi_k}{v_F} - \frac{\kappa}{2}\frac{\xi_k^2}{k_F v_F^2} + \frac{\zeta}{2}\frac{\xi_k^3}{k_F^2 v_F^3} + \mathcal{O}(\xi_k^4)\,.
$$

$$(2.1)$$

Here we assume the dispersions are rotational symmetric but non-parabolic. The non-parabolicity is encoded in the dimensionless parameters $\kappa$ and $\zeta$. In a Galilean invariant system, $\kappa = \zeta = 1$.

Since typical boson momentum $\boldsymbol{q}$ involved are much smaller than the Fermi wavevector $k_F$, we approximate the Jacobians and the density of states by

$$
J(|\boldsymbol{q}|,\cdot,\cdot) = \frac{1}{k_F|\boldsymbol{q}|}\,, \qquad \mathcal{N} = \frac{k_F}{2\pi v_F}\,.
$$

$$(2.2)$$

Here we have neglected higher order corrections in $|\boldsymbol{q}|/k_F$ or $\xi/(v_F|\boldsymbol{q}|)$. This approximation does not affect our main conclusions and it only provides subleading corrections to our calculation.

The first line of Eq.(17) of the main text, i.e. $L_m = L_m^{(0)} + L_m^{(1)} + L_m^{(2)} + \ldots$ are generated by expanding the Chebyshev polynomials in powers of $\xi$:

$$
\begin{aligned}
T_m\left(\frac{|\boldsymbol{k}|^2+|\boldsymbol{k}'|^2-|\boldsymbol{q}|^2}{2|\boldsymbol{k}||\boldsymbol{k}'|}\right) = -\cos\left(2m\phi_q\right) + \frac{m\left(\xi'+\xi\right)\cot\phi_q\sin\left(2m\phi_q\right)}{k_F v_F} \\
+ \frac{m\cot^3\phi_q}{4k_F^2 v_F^2}\left[\sin\left(2m\phi_q\right)\left[2(\kappa+1)\left(\xi^2+\xi'^2\right)-2(\kappa+2)\left(\xi^2+\xi'^2\right)\sec^2\phi_q+\left(\xi-\xi'\right)^2\sec^4\phi_q\right]\right. \\
\left. + 2m\left(\xi'+\xi\right)^2\tan\phi_q\cos\left(2m\phi_q\right)\right] + \mathcal{O}\left(\frac{\xi^3}{k_F^3 v_F^3}\right)\,,
\end{aligned}
$$

$$(2.3)$$

$$
\begin{aligned}
2T_m\left(\frac{|\boldsymbol{q}|^2+|\boldsymbol{k_1}|^2-|\boldsymbol{k}'|^2}{2|\boldsymbol{q}||\boldsymbol{k_1}|}\right)T_m\left(\frac{|\boldsymbol{q}|^2+|\boldsymbol{k_2}|^2-|\boldsymbol{k}''|^2}{2|\boldsymbol{q}||\boldsymbol{k_2}|}\right) = \cos(2m\phi_q)+1 \\
- \frac{m\csc(2\phi_q)\sin(2m\phi_q)\left(\xi''+\xi'+(\xi_1+\xi_2)\cos(2\phi_q)\right)}{k_F v_F} - \frac{m\sin^3\frac{\phi_q}{2}\cos\frac{\phi_q}{2}\csc^4\phi_q\sec^2\phi_q}{8\left(k_F^2 v_F^2\right)}\left[\right. \\
-\cos\phi_q\csc^2\frac{\phi_q}{2}\sin(2m\phi_q)\left(2\cos(2\phi_q)\left(-(\kappa-1)\left(\xi''^2+\xi'^2\right)+(\kappa+1)\xi_1^2+(\kappa+1)\xi_2^2\right)\right. \\
+ 2\kappa\left(\xi''^2+\xi'^2\right)-(\kappa+1)\left(\xi_1^2+\xi_2^2\right)\cos(4\phi_q)-\left((\kappa-1)\xi_1^2\right)-(\kappa-1)\xi_2^2+4\xi_2\xi''+4\xi_1\xi'\right) \\
+ 8m\cot\frac{\phi_q}{2}\cos^2(m\phi_q)\left(\xi''^2+\xi'^2+2\cos(2\phi_q)\left(\xi_2\xi''+\xi_1\xi'\right)+\left(\xi_1^2+\xi_2^2\right)\cos^2(2\phi_q)\right) \\
-16m\cot\frac{\phi_q}{2}\sin^2(m\phi_q)\left(\xi'+\xi_1\cos(2\phi_q)\right)\left(\xi''+\xi_2\cos(2\phi_q)\right)\right] + \mathcal{O}\left(\frac{\xi^3}{k_F^3 v_F^3}\right)\,.
\end{aligned}
$$

$$(2.4)$$

Here $\phi_q$ is defined as

$$\phi_q = \arccos \frac{|\boldsymbol{q}|}{2k_F} . \tag{2.5}$$

The vertical directions of Eq.(17) can then be generated by expanding (2.3) and (2.4) in powers of $|\boldsymbol{q}|^2$.

## III. DERIVATION OF EQ.(18)

At the order of $\delta_q^0 L_m^{(0)}$, only $L_{\text{MT+DOS}}$ contributes

$$\delta_q^0 L_m^{(0)}[F](i\omega, \xi) = g^2 \int \frac{d\omega'}{2\pi} \int \frac{\mathcal{N} d\xi'}{2\pi} \int |\boldsymbol{q}| d|\boldsymbol{q}| J(|\boldsymbol{k}|, |\boldsymbol{k}'|, |\boldsymbol{q}|) D(|\boldsymbol{q}|, i\omega - i\omega') \\ \times 2\left[iG(i\omega' + i\Omega/2, \xi') - iG(i\omega' - i\Omega/2, \xi')\right]\left[F(i\omega, \xi) - F(i\omega', \xi')\right] , \tag{3.1}$$

and obviously it is annihilated by the constant function $F(i\omega, \xi) = 1$.

To resolve the eigenvalue we apply first order perturbation theory

$$\lambda_m^{\text{even}} = \frac{\langle 1|\delta_q^1 L_m^{(0)}|1\rangle}{\langle 1|1\rangle} . \tag{3.2}$$

It is not hard to see that for even $m$, $L_{\text{AL,m}}^{(0)}[1] = 0$. This because at zeroth order in $\xi$, the integral over $\xi', \xi''$ and $\xi_2$ can be calculated with the Green's function factors in Eq.(1.19). The $\xi_2$ integral yields

$$\frac{\operatorname{sgn}(\omega_2 + \Omega/2) - \operatorname{sgn}(\omega_2 - \Omega/2)}{2} ,$$

which restricts $-\Omega/2 < \omega_2 < \Omega/2$. The $\xi'$ integral yields

$$\frac{\operatorname{sgn}(\omega_2 - \nu) + \operatorname{sgn}(\omega_2 + \nu)}{2} = \theta(|\omega_2| - |\nu|)\operatorname{sgn}\omega_2 .$$

This causes the integral to vanish because the input function $F(i\omega_2, \xi_2) = 1$ is even in $\omega_2$.

Therefore, we only need to consider first order $|\boldsymbol{q}|^2$ expansion of (1.15), which is

$$\delta_q^1 L_{\text{MT},m}^{(0)}[F](i\omega) = 2g^2 \mathcal{N} \int \frac{d\omega'}{2\pi} \int \frac{d|\boldsymbol{q}|}{k_F} D(|\boldsymbol{q}|, i\omega - i\omega') \frac{m^2|\boldsymbol{q}|^2}{2k_F^2} F(i\omega') . \tag{3.3}$$

Eq. (18) then follows from the fact that

$$\langle 1|1\rangle = \int \frac{d\omega}{2\pi} \int \frac{\mathcal{N} d\xi}{2\pi} \left[iG(i\omega + i\Omega/2, \xi) - iG(i\omega - i\Omega/2)\right] = \mathcal{N} \frac{\Omega}{2\pi} . \tag{3.4}$$

## IV. PROOF OF $L_m^{(0)}[1] = 0$ AND DERIVATION OF EQ.(20)

We first demonstrate that $L_m^{(0)}[1]$ to all order in $|\boldsymbol{q}|$ for odd $m$. To see this, we evaluate Eq.(1.19) on the constant function. The integral over $\omega_2, \xi_2$ and $\xi''$ can be evaluated to the boson self-energy due to rewriting of the Eliashberg equations below

$$\Pi(i\nu, |\boldsymbol{q}|) = -g^2 \mathcal{N}^2 (2\pi)^2 \int \frac{d\omega_2}{2\pi} \frac{d\xi_2}{2\pi} \frac{d\xi''}{2\pi} J(|\boldsymbol{k}|, |\boldsymbol{k}'|, |\boldsymbol{q}|) 2G(i\omega_2 + i\nu, \xi_2) G(i\omega_2, \xi'') . \tag{4.1}$$

Therefore, the result of $L^{(0)}_{\mathrm{AL},m}[1]$ becomes

$$
\begin{aligned}
L^{(0)}_{\mathrm{AL},m}[1] &= g^2 \mathcal{N} \int \frac{\mathrm{d}\omega'}{2\pi} \frac{\mathrm{d}\xi'}{2\pi} \int |\boldsymbol{q}| \mathrm{d}|\boldsymbol{q}| J(|\boldsymbol{k}|_1, |\boldsymbol{k}'|, |\boldsymbol{q}|) D(|\boldsymbol{q}|, i\omega - i\omega') \\
&\quad \times 2[iG(i\omega' + i\Omega/2, \xi') - iG(i\omega' - i\Omega/2, \xi')] \times (-2) T^2_m \left( \frac{|\boldsymbol{q}|}{2k_F} \right)
\end{aligned}
\tag{4.2}
$$

Combining with Eq.(1.15), we obtain

$$
\begin{aligned}
L^{(0)}_m[1](i\omega, \xi) &= g^2 \mathcal{N} \int \frac{\mathrm{d}\omega'}{2\pi} \frac{\mathrm{d}\xi'}{2\pi} \int |\boldsymbol{q}| \mathrm{d}|\boldsymbol{q}| J(|\boldsymbol{k}|, |\boldsymbol{k}'|, |\boldsymbol{q}|) D(|\boldsymbol{q}|, i\omega - i\omega') \\
&\quad \times 2[iG(i\omega' + i\Omega/2, \xi') - iG(i\omega' - i\Omega/2, \xi')] \left[ 1 - T_m \left( 1 - \frac{|\boldsymbol{q}|^2}{2k_F} \right) - 2T^2_m \left( \frac{|\boldsymbol{q}|}{2k_F} \right) \right] = 0 \,.
\end{aligned}
\tag{4.3}
$$

As a consequence of trigonometric identity, the term in the bracket vanishes identically.

The vanishing of $L^{(0)}_m[1]$ is related to the fact that on a 2D convex Fermi surface, the two-to-two scattering configurations can only be forward scattering, particle exchange or head-on scattering, and neither of these configurations can relax the odd angular harmonics of the Fermi surface deformation.

Noticing the fact that $G(i\omega, \xi) = -G(-i\omega, -\xi)$, the kinetic operators $L_m$ has a particle-hole symmetry under which

$$
L^{(n)}_m \to (-1)^n L^{(n)}_m \,.
$$

Therefore, first order perturbation in $\xi$ $\langle 1|L^{(1)}_m|1\rangle$ vanishes identically and we have to consider the second order expression

$$
\lambda^{\mathrm{odd}}_m = \frac{\delta^1_q}{\langle 1|1\rangle} \left[ \langle 1|L^{(2)}_m|1\rangle - \langle 1|L^{(1)}_m \frac{1}{L^{(0)}_m} L^{(1)}_m|1\rangle \right] \,.
\tag{4.4}
$$

The hard part of evaluating (4.4) is to perform the functional inverse of $L^{(0)}_m$. We first compute $L^{(1)}_m |1\rangle$. Using manipulations similar to (4.1) and (4.2), we obtain

$$
\begin{aligned}
L^{(1)}_m[1](i\omega, \xi) &= g^2 \mathcal{N} \int \frac{\mathrm{d}\omega'}{2\pi} \frac{\mathrm{d}\xi'}{2\pi} \int |\boldsymbol{q}| \mathrm{d}|\boldsymbol{q}| J(|\boldsymbol{k}|, |\boldsymbol{k}'|, |\boldsymbol{q}|) D(|\boldsymbol{q}|, i\omega - i\omega') \\
&\quad \times 2[iG(i\omega' + i\Omega/2, \xi') - iG(i\omega' - i\Omega/2, \xi')] \frac{-m}{k_F v_F} \left[ \xi \sin(2m\phi_q) \csc(2\phi_q) + \xi'(\cot\phi_q - \csc(2\phi_q)) \sin(2m\phi_q) \right] \,.
\end{aligned}
\tag{4.5}
$$

We compare this with $L^{(0)}_m |\xi/(k_F v_F)\rangle$, which is

$$
\begin{aligned}
L^{(0)}_m \left[ \frac{\xi}{k_F v_F} \right] (i\omega, \xi) &= g^2 \mathcal{N} \int \frac{\mathrm{d}\omega'}{2\pi} \frac{\mathrm{d}\xi'}{2\pi} \int |\boldsymbol{q}| \mathrm{d}|\boldsymbol{q}| J(|\boldsymbol{k}|, |\boldsymbol{k}'|, |\boldsymbol{q}|) D(|\boldsymbol{q}|, i\omega - i\omega') \\
&\quad \times 2[iG(i\omega' + i\Omega/2, \xi') - iG(i\omega' - i\Omega/2, \xi')] \left[ \xi + \xi' \cos(2m\phi_q) \right] \,.
\end{aligned}
\tag{4.6}
$$

We are interested in the case of small-angle scattering, so we expand Eqs.(4.5) and (4.6) to order $|\boldsymbol{q}|^2$, yielding

$$
\begin{aligned}
(\delta^0_q + \delta^1_q) L^{(1)}_m[1](i\omega, \xi) &= g^2 \mathcal{N} \int \frac{\mathrm{d}\omega'}{2\pi} \frac{\mathrm{d}\xi'}{2\pi} \int |\boldsymbol{q}| \mathrm{d}|\boldsymbol{q}| J(|\boldsymbol{k}|, |\boldsymbol{k}'|, |\boldsymbol{q}|) D(|\boldsymbol{q}|, i\omega - i\omega') \\
&\quad \times 2[iG(i\omega' + i\Omega/2, \xi') - iG(i\omega' - i\Omega/2, \xi')](-m^2) \left[ \left( 1 - \frac{(m^2-1)|\boldsymbol{q}|^2}{6k_F^2} \right) \frac{\xi}{k_F v_F} - \left( 1 - \frac{(m^2+2)|\boldsymbol{q}|^2}{6k_F^2} \right) \frac{\xi'}{k_F v_F} \right] \,.
\end{aligned}
\tag{4.7}
$$

$$
\begin{aligned}
(\delta^0_q + \delta^1_q) L^{(0)}_m \left[ \frac{\xi}{k_F v_F} \right] (i\omega, \xi) &= g^2 \mathcal{N} \int \frac{\mathrm{d}\omega'}{2\pi} \frac{\mathrm{d}\xi'}{2\pi} \int |\boldsymbol{q}| \mathrm{d}|\boldsymbol{q}| J(|\boldsymbol{k}|, |\boldsymbol{k}'|, |\boldsymbol{q}|) D(|\boldsymbol{q}|, i\omega - i\omega') \\
&\quad \times 2[iG(i\omega' + i\Omega/2, \xi') - iG(i\omega' - i\Omega/2, \xi')] \left[ \frac{\xi}{k_F v_F} - \left( 1 - \frac{m^2 |\boldsymbol{q}|^2}{2k_F^2} \right) \frac{\xi'}{k_F v_F} \right] \,.
\end{aligned}
\tag{4.8}
$$

Therefore, we have (we drop $\delta_q^0 + \delta_q^1$ for clarity)

$$\frac{1}{L_m^{(0)}} L_m^{(1)} |1\rangle = -m^2 |\frac{\xi}{k_F v_F}\rangle + |h\rangle \ , \tag{4.9}$$

where the correction $|h\rangle$ is of order $|\boldsymbol{q}|^2$.

We rewrite Eq.(4.9) as

$$L_m^{(0)} |h\rangle = |b\rangle \equiv m^2 L_m^{(0)} |\frac{\xi}{k_F v_F}\rangle + L_m^{(1)} |1\rangle \ . \tag{4.10}$$

The explicit expression for function $|b\rangle$ is

$$b(i\omega, \xi) = g^2 \mathcal{N} \int \frac{\mathrm{d}\omega'}{2\pi} \frac{\mathrm{d}\xi'}{2\pi} \int |\boldsymbol{q}| \mathrm{d}|\boldsymbol{q}| J(|\boldsymbol{k}|, |\boldsymbol{k}'|, |\boldsymbol{q}|) D(|\boldsymbol{q}|, i\omega - i\omega')$$
$$\times 2[iG(i\omega' + i\Omega/2, \xi') - iG(i\omega' - i\Omega/2, \xi')] \frac{m^2(m^2-1)|\boldsymbol{q}|^2}{6k_F^2} \left[ \frac{\xi}{k_F v_F} + 2\frac{\xi'}{k_F v_F} \right] \ . \tag{4.11}$$

Therefore, the eigenvalue is

$$\lambda_m^{\mathrm{odd}} = \frac{1}{\langle 1|1\rangle} \delta_q^1 \left[ \langle 1|L_m^{(2)}|1\rangle + m^2 \langle 1|L_m^{(1)}| \frac{\xi}{k_F v_F}\rangle - \langle 1|L_m^{(1)}|h\rangle \right] \tag{4.12}$$

The last term of Eq.(4.12) can be computed as follows. Since $|h\rangle$ is first order in $|\boldsymbol{q}|^2$, everything else only needs to be computed to zeroth order in $|\boldsymbol{q}|^2$. We therefore utilize (4.9) to zeroth order in $|\boldsymbol{q}|$, which states that

$$\delta_q^0 L_m^{(1)} |1\rangle = -m^2 \delta_q^0 L_m^{(0)} |\frac{\xi}{k_F v_F}\rangle \ . $$

Combining with Eq.(4.10), we therefore obtain

$$-\delta_q^1 \langle 1|L_m^{(1)}|h\rangle = m^2 \langle \frac{\xi}{k_F v_F}|\delta_q^0 L_m^{(0)}|h\rangle = m^2 \langle \frac{\xi}{k_F v_F}|b\rangle \ . \tag{4.13}$$

So we have

$$\lambda_m^{\mathrm{odd}} = \frac{1}{\langle 1|1\rangle} \left[ \langle 1|\delta_q^1 L_m^{(2)}|1\rangle + m^2 \langle 1|\delta_q^1 L_m^{(1)}| \frac{\xi}{k_F v_F}\rangle + m^2 \langle \frac{\xi}{k_F v_F}|b\rangle \right] \tag{4.14}$$

Substituting the explicit expansions, we obtain Eq.(20) of the main text:

$$\lambda_m^{\mathrm{odd}} = \frac{2g^2 \mathcal{N}}{\Omega/(2\pi)} \int_{-\Omega/2}^{\Omega/2} \frac{\mathrm{d}\omega \mathrm{d}\omega'}{(2\pi)^2} \int_{-\infty}^{\infty} \frac{\mathrm{d}\xi \mathrm{d}\xi'}{(2\pi)^2} \int_0^{\infty} \frac{\mathrm{d}|\boldsymbol{q}|}{k_F}$$
$$\times D(|\boldsymbol{q}|, i\omega - i\omega') [iG(i\omega + i\Omega/2, \xi) - iG(i\omega - i\Omega/2, \xi)]$$
$$\times [iG(i\omega' + i\Omega/2, \xi') - iG(i\omega' - i\Omega/2, \xi')] \times \frac{|\boldsymbol{q}|^2}{k_F^2} \frac{m^2(m^2-1)^2(\xi+\xi')^2}{8k_F^2 v_F^2} \ . \tag{4.15}$$

## V. SOME DETAILS IN THE EVALUATION OF EQS.(22),(23),(25),(26),(27)

Here we provide some details in applying the our results to the Ising-Nematic QCP.

In evaluating both $\lambda_m^{\mathrm{even}}$ and $\lambda_m^{\mathrm{odd}}$, we need to compute the integral

$$\int_0^{\infty} \mathrm{d}|\boldsymbol{q}| D(|\boldsymbol{q}|, i\Omega)|\boldsymbol{q}|^2 \ . \tag{5.1}$$

Since the integral might not converge in the UV, we apply dimensional regularization by evaluating $\int d|\boldsymbol{q}| D(|\boldsymbol{q}|, i\Omega) |\boldsymbol{q}|^\eta$ and then continue to $\eta \to 2$.

In the FL regime, $D(|\boldsymbol{q}|, i\Omega)$ is

$$D(|\boldsymbol{q}|, i\Omega) = \frac{1}{|\boldsymbol{q}|^2 + m_b^2 + \gamma \frac{|\Omega|}{|\boldsymbol{q}|}}. \tag{5.2}$$

We evaluate it in the limit of $m_b \to \infty$, and we find

$$\int_0^\infty d|\boldsymbol{q}| D(|\boldsymbol{q}|, i\Omega) |\boldsymbol{q}|^2 = -m_b \left( \frac{\pi}{2} + \frac{|\Omega|}{\omega_{\mathrm{FL}}} \right), \tag{5.3}$$

where $\omega_{\mathrm{FL}} = m_b^3/\gamma$.

In the NFL regime (perturbative or not), $m_b^2 = 0$, and we have

$$\int_0^\infty d|\boldsymbol{q}| D(|\boldsymbol{q}|, i\Omega) |\boldsymbol{q}|^2 = -\frac{2\pi}{3\sqrt{3}} \gamma^{1/3} |\Omega|^{1/3}. \tag{5.4}$$

Combining Eqs. (5.2) and (5.4) with Eq.(18) of the main text, we can obtain Eqs. (22) and (25).

Next we discuss how Eq. (20) is evaluated. We write the Green's function as $G(i\omega, \xi) = 1/(A(i\omega) - \xi)$ where $A(i\omega) = i\omega - \Sigma(i\omega)$.

We evaluate the integral by computing the $\xi$ and $\xi'$ integrations with the contour method:

$$\begin{aligned}
\lambda_m^{\mathrm{odd}} &= \frac{2g^2 \mathcal{N}}{\Omega/(2\pi)} \int_{-\Omega/2}^{\Omega/2} \frac{d\omega d\omega'}{(2\pi)^2} \int_0^\infty \frac{d|\boldsymbol{q}|}{k_F} D(|\boldsymbol{q}|, i\omega - i\omega') \\
&\times \frac{|\boldsymbol{q}|^2}{k_F^2} \frac{m^2(m^2-1)^2}{4k_F^2 v_F^2} \left[ \frac{A_+^2 + A_-^2}{2} + \frac{A_+ + A_-}{2} \frac{A_+' + A_-'}{2} \right].
\end{aligned} \tag{5.5}$$

Here $A_\pm = A(i\omega \pm i\Omega/2)$ and $A_\pm' = A(i\omega' \pm i\Omega/2)$. Recall that in the retarded branch $\Omega > 0$ and $-\Omega/2 < \omega < \Omega/2$.

In the FL regime, we can set $A(i\omega) = i\omega(1 + \pi c_f')$ where the $(1 + \pi c_f')$ factor reflects the renormalization of quasiparticle residue. In the perturbative NFL regime, we use $A(i\omega) = i\omega$. In the NFL regime, we use $A(i\omega) = i \mathrm{sgn}\,\omega c_f |\omega|^{2/3}$. Substitute Eqs.(5.2) and (5.4), and evaluate the $\omega, \omega'$ integrals, we obtain Eqs.(23),(26),(27) of the main text.

## VI. EFFECTS OF THERMAL FLUCTUATION

In this part we show that thermal fluctuation does not lead to the instability. When thermal fluctuations are included, the boson acquires a thermal mass $\Delta = \Delta(T)$. We only consider the thermal effects when boson carries zero Matsubara frequency, so the thermal boson propagator is

$$D_{\mathrm{th}}(i\Omega, \boldsymbol{q}) = \frac{1}{|\boldsymbol{q}|^2 + \Delta^2} 2\pi T \delta(\Omega). \tag{6.1}$$

Associated with that, the fermion has a self energy $\Sigma_T(i\omega) = -i\mathrm{sgn}\,\omega \times \Gamma(T)$, where $\Gamma(T) \sim T/\Delta(T)$. Substitute these into Eq.(5.5), we obtain

$$\lambda_m^{\mathrm{odd}}(i\Omega) = \frac{2g^2 \mathcal{N}}{k_F} \frac{T}{\Delta} \times \frac{\pi \Delta^2}{2k_F^2} \times \frac{m^2(m^2-1)^2 \Gamma^2}{4(k_F v_F)^2}, \tag{6.2}$$

which is $\Omega$-independent and positive. For the Ising-Nematic QCP, $\Delta(T)^2 \sim T \ln(1/T)$ [1, 2].


\fi
\end{document}